\documentclass[aps,prd,superscriptaddress,twoside,twocolumn,nofootinbib,showpacs,floatfix]{revtex4-1}
\usepackage{amsmath,amssymb}
\usepackage{graphicx,bm,braket}
\usepackage{subfigure}
\usepackage{color}
\pdfpagewidth=\paperwidth
\pdfpageheight=\paperheight
\allowdisplaybreaks
\usepackage{epstopdf}

\newcommand*{\hc}{\text{H.\,c.}}

\begin{document}

\title{\boldmath Effects of $N(2080){3/2}^-$ and $N(2270)3/2^-$ molecules on $K^\ast \Sigma$ photoproduction}

\author{Di Ben}
\affiliation{School of Nuclear Science and Technology, University of Chinese Academy of Sciences, Beijing 101408, China}
\affiliation{CAS Key Laboratory of Theoretical Physics, Institute of Theoretical Physics, Chinese Academy of Sciences, Beijing 100190, China}

\author{Ai-Chao Wang}
\affiliation{College of Science, China University of Petroleum (East China), Qingdao 266580, China}

\author{Fei Huang}
\email{huangfei@ucas.ac.cn}
\affiliation{School of Nuclear Science and Technology, University of Chinese Academy of Sciences, Beijing 101408, China}

\author{Bing-Song Zou}
\email{zoubs@itp.ac.cn}
\affiliation{CAS Key Laboratory of Theoretical Physics, Institute of Theoretical Physics, Chinese Academy of Sciences, Beijing 100190, China}
\affiliation{School of Physical Sciences, University of Chinese Academy of Sciences, Beijing 100049, China}
\affiliation{Southern Center for Nuclear-Science Theory (SCNT), Institute of Modern Physics, Chinese Academy of Sciences, Huizhou 516000, China}

\date{\today} 

\begin{abstract}
In our previous work [Phys. Rev. C {\bf 98}, 045209 (2018)], the available differential cross-section data for $\gamma p\to K^{\ast +}\Sigma^0$ and $\gamma p \to K^{\ast 0}\Sigma^+$ have been analyzed within an effective Lagrangian approach. It was found that one needs to introduce the $s$-channel $\Delta(1905){5/2}^+$ resonance exchange besides the $t$-channel $K$, $\kappa$, and $K^\ast$ exchanges, the $s$-channel $N$ and $\Delta$ exchanges, the $u$-channel $\Lambda$, $\Sigma$, and $\Sigma^\ast$ exchanges, and the generalized contact term in constructing the reaction amplitudes to describe the data. In the present work, we re-analyze the available data for $\gamma p\to K^{\ast +}\Sigma^0$ and $\gamma p \to K^{\ast 0}\Sigma^+$ by considering the contributions from the $N(2080){3/2}^-$ and $N(2270)3/2^-$ molecules instead of any nucleon resonances in the $s$ channel, where the $N(2080)3/2^-$ was proposed to be a $K^\ast \Sigma$ molecule as the strange partner of the $P_c^+(4457)$ hadronic molecular state, and the $N(2270)3/2^-$ was assumed to be a $K^*\Sigma^*$ molecule as the strange partner of the $\bar{D}^\ast \Sigma^\ast_c$ bound states that are predicated as members in the same heavy-quark spin symmetry multiplet as the $P_c$ states. It turns out that all the available cross-section data can be well reproduced, indicating that the molecular structures of the possible $N(2080){3/2}^-$ and $N(2270)3/2^-$ states are compatible with the available data for $K^\ast\Sigma$ photoproduction reactions. Further analysis shows that for both $\gamma p\to K^{\ast +}\Sigma^0$ and $\gamma p \to K^{\ast 0}\Sigma^+$ reactions, the $N(2080){3/2}^-$ exchange provides dominant contributions to the cross-sections in the near-threshold energy region, and significant contributions from the $N(2270)3/2^-$ exchange to the cross-sections in the higher energy region are also found. Predictions of the beam asymmetry $\Sigma$, target asymmetry $T$, and recoil baryon asymmetry $P$ are presented and compared with those from our previous work. Measurements of the data on these observables are called on to further constrain the reaction mechanisms of $K^\ast\Sigma$ photoproduction reactions and to verify the molecular scenario of the $N(2080){3/2}^-$ and $N(2270)3/2^-$ states.
\end{abstract}



\maketitle

\section{Introduction}\label{Sec:intro}

Multiquark states that are beyond the traditional quark-antiquark ($q\bar{q}$) mesons and three-quark ($qqq$) baryons have been one of the most interested topics in hadron physics from the dawn of the quark model. In the past few decades, although a lot of multiquark states have been theoretically predicated or experimentally reported, no compelling multiquark candidates were unambiguously identified until 2015 when the LHCb Collaboration presented striking evidence for $J/\psi \, p$ resonances, named as $P_c^+(4380)$ and $P_c^+(4450)$, in $\Lambda^0_b\to K^- J/\psi \, p$ decays \cite{Aaij:2015tga}. In 2019, the LHCb Collaboration further reported the $P_c^+(4312)$ state and a two-peak structure of the $P_c^+(4450)$ state which is resolved into $P_c^+(4440)$ and $P_c^+(4457)$ \cite{Aaij:2019}. Unlike the low-energy nucleon resonances whose excitation energies are hundreds of MeV and thus can be accommodated as either excited three-quark states or baryon-meson states or compact pentaquark states, the $P_c$ states have more than $3$ GeV excitation energies, definitely excluding the possibility of being excited three-quark configuration dominated states. Indeed, they are the most promising candidates for hidden-charm pentaquark states or baryon-meson states as predicated in Refs.~\cite{Wu:2010jy,Wu:2010vk,Wang:2011prc,Yang:2011wz,Yuan:2012wz,Xiao:2013yca}.

In literature, there are many theoretical investigations on the nature of the $P_c$ states \cite{Guo:2017jvc,Chen:2016qju}. The fact that the reported masses of $P_c^+(4380)$ and $P_c^+(4457)$ locate just below the thresholds of $\bar{D}\Sigma_c^\ast$ and $\bar{D}^\ast\Sigma_c$ at $4382$ MeV and $4459$ MeV seems strongly support the interpretation of $P_c^+(4380)$ and $P_c^+(4457)$ as hadronic molecules composed of $\bar{D} \Sigma_c^\ast$ and $\bar{D}^\ast \Sigma_c$, respectively. Analogously, in the light quark sector, as the masses of $N(1875){3/2}^-$ and $N(2080){3/2}^-$ are just below the thresholds of $K\Sigma^\ast$ and $K^\ast\Sigma$ at $1880$ MeV and $2086$ MeV, respectively, the $N(1875){3/2}^-$ and $N(2080){3/2}^-$ are proposed to be the strange partners of the $P_c^+(4380)$ and $P_c^+(4457)$ molecular states \cite{He:2017aps,Lin:2018kcc}. In Ref.~\cite{Lin:2018kcc}, the decay patterns of $N(1875){3/2}^-$ and $N(2080){3/2}^-$ as $S$-wave $K\Sigma^\ast$ and $K^\ast\Sigma$ molecular states were calculated within an effective Lagrangian approach, and it was found that the measured decay properties of $N(1875){3/2}^-$ and $N(2080){3/2}^-$ can be reproduced well, supporting the molecule interpretation of the $N(1875){3/2}^-$ and $N(2080){3/2}^-$ states.

In addition to the $\bar{D} \Sigma_c^\ast$ and $\bar{D}^\ast \Sigma_c$ molecules, the $\bar{D}^\ast \Sigma^\ast_c$ bound states were also predicted as members of heavy-quark spin symmetry multiplet in study of the $P_c$ states in a contact-range effective field theory \cite{Liu:2019}. The analogue of $\bar{D}^\ast \Sigma^\ast_c$ bound states in the light quark sector is called $N(2270)$ with spin-parity ${1/2}^-$ or ${3/2}^-$ \cite{Wu:2023}, which is just below the threshold of $K^\ast \Sigma^\ast$ at $2277$ MeV.

The limited number of available data points for $\bar{D}\Sigma_c^\ast$ and $\bar{D}^\ast\Sigma_c$ interactions restrains, to some extent, our exploration for the nature of $P_c^+(4380)$ and $P_c^+(4457)$ states. In contrast, the situation in the light quark section is much better. So far, lots of experimental data points on differential and total cross sections for $K\Sigma^\ast $ and $K^\ast\Sigma$ photoproductions are available \cite{Hleiqawi:2005sz,Hleiqawi:2007ad,Nanova:2008kr,Hwang2012,Wei:2013,Moriya:2013}, providing good opportunities to investigate the possible molecular scenario of the $N(1875){3/2}^-$ and $N(2080){3/2}^-$ states. In Ref.~\cite{Wu:2023}, the $\gamma p\to \phi p$ reaction has been studied in an effective Lagrangian approach, and it is shown that the available data for this reaction can be well described by considering the $N(2080){3/2}^-$ and $N(2270){3/2}^-$ molecules, while the $N(2270)$ with spin-parity ${1/2}^-$ is not favored by the $\phi$ photoproduction data.
In the present work, we focus on the $\gamma p\to K^{\ast +}\Sigma^0$ and $\gamma p \to K^{\ast 0}\Sigma^+$ reactions to test the effects of $N(2080){3/2}^-$ as $K^\ast \Sigma$ molecule and $N(2270){3/2}^-$ as $K^\ast \Sigma^\ast$ molecule on $K^\ast \Sigma$ photoproduction reactions.

Note that in the most recent Particle Data Group (PDG) review \cite{PDG2022}, the two-star $N(2080){3/2}^-$ listed before the 2012 review has been split into two three-star states, i.e. the $N(1875){3/2}^-$ and $N(2120){3/2}^-$ states. For $N(1875){3/2}^-$, the Breit-Wigner mass and width are claimed to be $1850 < W < 1920$ MeV and $120 < \Gamma < 250$ MeV, respectively. For $N(2120){3/2}^-$, the corresponding values are $2060 < W < 2160$ MeV and $260 < \Gamma < 360$ MeV, respectively. Since in general the hadronic molecules are very shallowly bounded, in Ref.~\cite{Lin:2018kcc} the masses of $N(1875){3/2}^-$ and $N(2120){3/2}^-$ were taken as $1875$ MeV and $2080$ MeV, respectively, and the old name $N(2080){3/2}^-$ was used for the $N(2120){3/2}^-$ state. In the present work, we follow Ref.~\cite{Lin:2018kcc} to use the name $N(2080){3/2}^-$ for the possible $K^\ast \Sigma$ molecule. We mention that this molecule is not necessarily to be the $N(2120){3/2}^-$ resonance in PDG review \cite{PDG2022}.

The $K^\ast \Sigma$ photoproduction process has ever been investigated in several theoretical works by use of either chiral quark model \cite{Zhao:2001jw} or effective Lagrangian approaches \cite{Oh:2006in,Kim:20132,Wang:2018vlv}. Our previous work of Ref.~\cite{Wang:2018vlv} provides so far the most recent and most comprehensive analysis of the available data for $\gamma p\to K^{\ast +}\Sigma^0$ and $\gamma p \to K^{\ast 0}\Sigma^+$ reactions. In Ref.~\cite{Wang:2018vlv}, it was found that the $K^\ast \Sigma$ photoproduction data can be well reproduced by introducing the $s$-channel $\Delta(1905)5/2^+$ resonance exchange in addition to the $t$-channel $K$, $\kappa$, $K^\ast$ exchanges, $s$-channel nucleon and $\Delta$ exchanges, $u$-channel $\Lambda$, $\Sigma$, $\Sigma^\ast$ exchanges, and generalized contact term in constructing the reaction amplitudes. The $\Delta(1905)5/2^+$ resonance exchange was found to dominate the cross sections of $\gamma p\to K^{\ast +}\Sigma^0$ and provide considerable contributions to the cross sections of $\gamma p \to K^{\ast 0}\Sigma^+$ near the threshold energy region.

In the present work, we re-analyze the data for $\gamma p\to K^{\ast +}\Sigma^0$ and $\gamma p \to K^{\ast 0}\Sigma^+$ within the effective Lagrangian approach as employed in Ref.~\cite{Wang:2018vlv}. Our purpose is to investigate the effects of $N(2080){3/2}^-$ as $K^\ast \Sigma$ molecular state and $N(2270)3/2^-$ as $K^*\Sigma^*$ molecular state on $K^\ast \Sigma$ photoproduction reactions. Instead of introducing in $s$ channel the $\Delta(1905)5/2^+$ resonance exchange as done in Ref.~\cite{Wang:2018vlv}, we now consider the contributions from the $N(2080){3/2}^-$ and $N(2270)3/2^-$ molecules in addition to the background contributions, i.e., the contributions from all diagrams other than the $\Delta(1905)5/2^+$ resonance exchange considered in Ref.~\cite{Wang:2018vlv}. Our results show that all the available data for $\gamma p\to K^{\ast +}\Sigma^0$ and $\gamma p \to K^{\ast 0}\Sigma^+$ can be well described in the energy region considered, indicating that the $K^\ast \Sigma$ molecular picture of $N(2080){3/2}^-$ and the $K^\ast \Sigma^\ast$ molecular picture of $N(2270)3/2^-$ are compatible with the available data of $K^\ast \Sigma$ photoproduction reactions. The individual contributions of the $N(2080){3/2}^-$ and $N(2270)3/2^-$ molecules to the cross sections are discussed. The reaction mechanisms are analyzed and compared with those extracted from Ref.~\cite{Wang:2018vlv}. The predictions of the beam asymmetry $\Sigma$, target asymmetry $T$, and recoil baryon asymmetry $P$ that can distinguish the reaction models constructed in the present work and Ref.~\cite{Wang:2018vlv} are presented for future experiments.

 The paper is organized as follows. In Sec.~\ref{Sec:forma}, we briefly introduce the framework of our theoretical model. In Sec.~\ref{Sec:results}, the results of our theoretical calculations with some discussions are presented. Finally, we give a brief summary and conclusions in Sec.~\ref{sec:summary}.

\section{Formalism}\label{Sec:forma}

\begin{figure}[tb]
    \centering
    {\vglue 0.15cm}
    \subfigure[~$s$ channel]{
    \includegraphics[width=0.45\columnwidth]{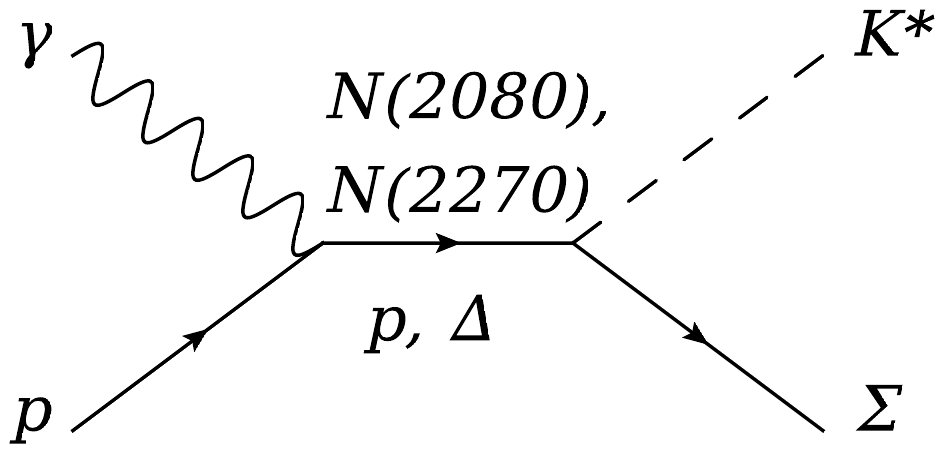}}  {\hglue 0.4cm}
    \subfigure[~$t$ channel]{
    \includegraphics[width=0.45\columnwidth]{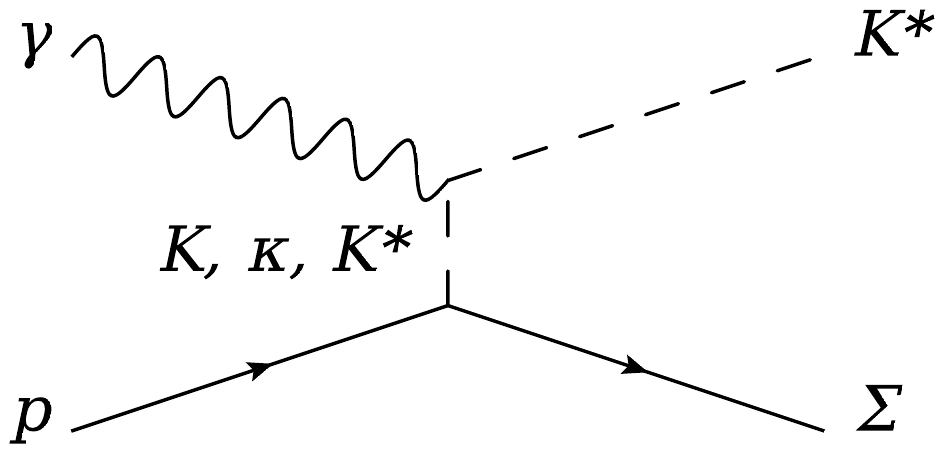}} \\[6pt]
    \subfigure[~$u$ channel]{
    \includegraphics[width=0.45\columnwidth]{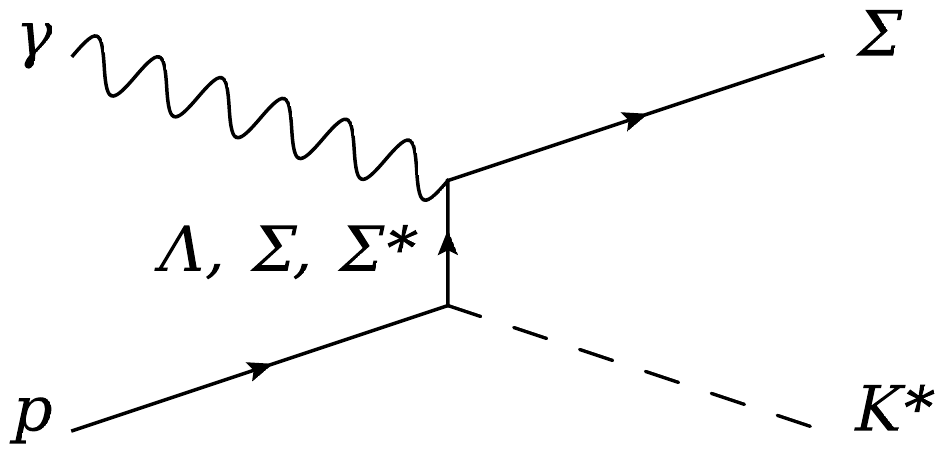}} {\hglue 0.4cm}
    \subfigure[~Interaction current]{
    \includegraphics[width=0.45\columnwidth]{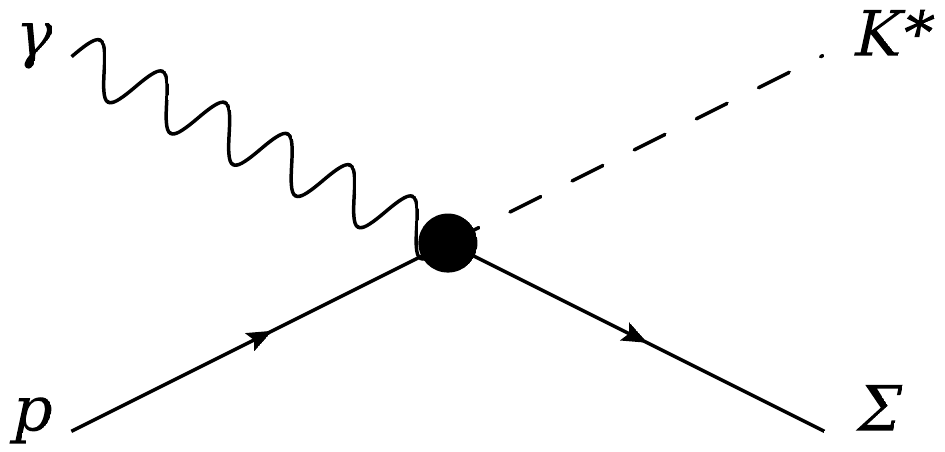}}
    \caption{
    Generic structure of the $K^\ast$ photoproduction amplitude for $\gamma p \rightarrow K^\ast \Sigma$. Time proceeds from left to right.}
    \label{fig1}
\end{figure}

In effective Lagrangian approach, the amplitude of $K^\ast\Sigma$ photoproduction process can be expressed as
\begin{equation}
    \mathcal{M} = \mathcal{M}_s + \mathcal{M}_t + \mathcal{M}_u + \mathcal{M}_{\rm int},
    \label{1}
\end{equation}
where $\mathcal{M}_s$, $\mathcal{M}_t$, and $\mathcal{M}_u$ denote the amplitudes obtained straightforwardly from the $s$-, $t$-, and $u$-channel tree-level Feynman diagrams, respectively, with $s$, $t$, and $u$ being the Mandelstam variables of the internally exchanged particles. The last term $\mathcal{M}_{\rm int}$ is the interaction current arising from the photon attaching to the internal structure of the $\Sigma N K^\ast$ interaction vertex. All these four terms in Eq.~\eqref{1} are diagrammatically
depicted in Fig.~\ref{fig1}.

As shown in Fig.~\ref{fig1}, the following contributions are considered in the present work: (i) $N$, $\Delta$, $N(2080){3/2}^-$, and $N(2270)3/2^-$ exchanges in the $s$ channel, (ii) $K$, $\kappa$, and $K^\ast$ exchanges in the $t$ channel,  (iii) $\Sigma$, $\Lambda$, and $\Sigma^\ast$ exchanges in the $u$ channel, and (iv) the interaction current. The most parts of the formalism including the Lagrangians, propagators, form factors attached to hadronic vertices, the gauge-invariance preserving term, and the interaction coupling constants are referred to Ref.~\cite{Wang:2018vlv}. For the simplicity of the present paper, we do not repeat them here. In the following subsections, we just present the additional parts of the theoretical formalism.

\subsection{Lagrangians and couplings for $N(2080){3/2}^-$ and $N(2270)3/2^-$} \label{Sec:2080}

The $N(2080){3/2}^-$ which is treated as a bound state of $K^\ast$ and $\Sigma$, and the $N(2270)3/2^-$ which is treated as a bound state of $K^*$ and $\Sigma^*$,  are considered in the present work to construct the $s$-channel reaction amplitude. The effective Lagrangians for $N(2080){3/2}^-$ and $N(2270)3/2^-$ coupled with $K^\ast \Sigma$ read
\begin{equation}
    \mathcal{L}_{K^\ast\Sigma R}^{3/2^-} = g_{K^\ast\Sigma R} {\bar R}_\mu \Sigma K^{\ast\mu} + \hc,
    \label{hms}
\end{equation}
where $R \equiv N(2080){3/2}^-$ or $N(2270)3/2^-$.

Considering that the $N(2080){3/2}^-$ is assumed to be a pure $S$-wave molecular state of $K^\ast$ and $\Sigma$, the coupling constant $g_{K^\ast\Sigma R}$ can be estimated model-independently
with the Weinberg compositeness criterion, which gives \cite{Lin:2017mtz,Baru:2003qq,Weinberg:1965zz}
\begin{equation}
    g_{K^\ast\Sigma R}^2 = \frac{4\pi}{4 M_R M_\Sigma}  \frac{\left(M_{K^\ast}+M_\Sigma\right)^{5/2}} {\left(M_{K^\ast} M_\Sigma\right)^{1/2}} \sqrt{32\,\epsilon},   \label{eq:coupling}
\end{equation}
where $M_R$, $M_{K^\ast}$, and $M_\Sigma$ denote the masses of $N(2080){3/2}^-$, $K^\ast$, and $\Sigma$, respectively, and $\epsilon$ is the $K^\ast\Sigma$ binding energy,
\begin{equation}
    \epsilon \equiv M_{K^\ast} + M_\Sigma - M_R.
    \label{mass}
\end{equation}
Following Ref.~\cite{Lin:2018kcc}, we take the mass of $N(2080){3/2}^-$ to be $M_R=2080$ MeV. Then one gets from Eq.~\eqref{eq:coupling}
\begin{equation}
g_{K^\ast\Sigma R} = 1.72.
\end{equation}
Note that in practical calculations, the isospin factors $\sqrt{2/3}$ and $\sqrt{1/3}$ are multiplied to the $N(2080)\Sigma^+ K^{\ast 0}$ and $N(2080)\Sigma^0 K^{\ast +}$ vertices, respectively.

\begin{figure}[tb]
    \centering
    \includegraphics[width=0.27\textwidth]{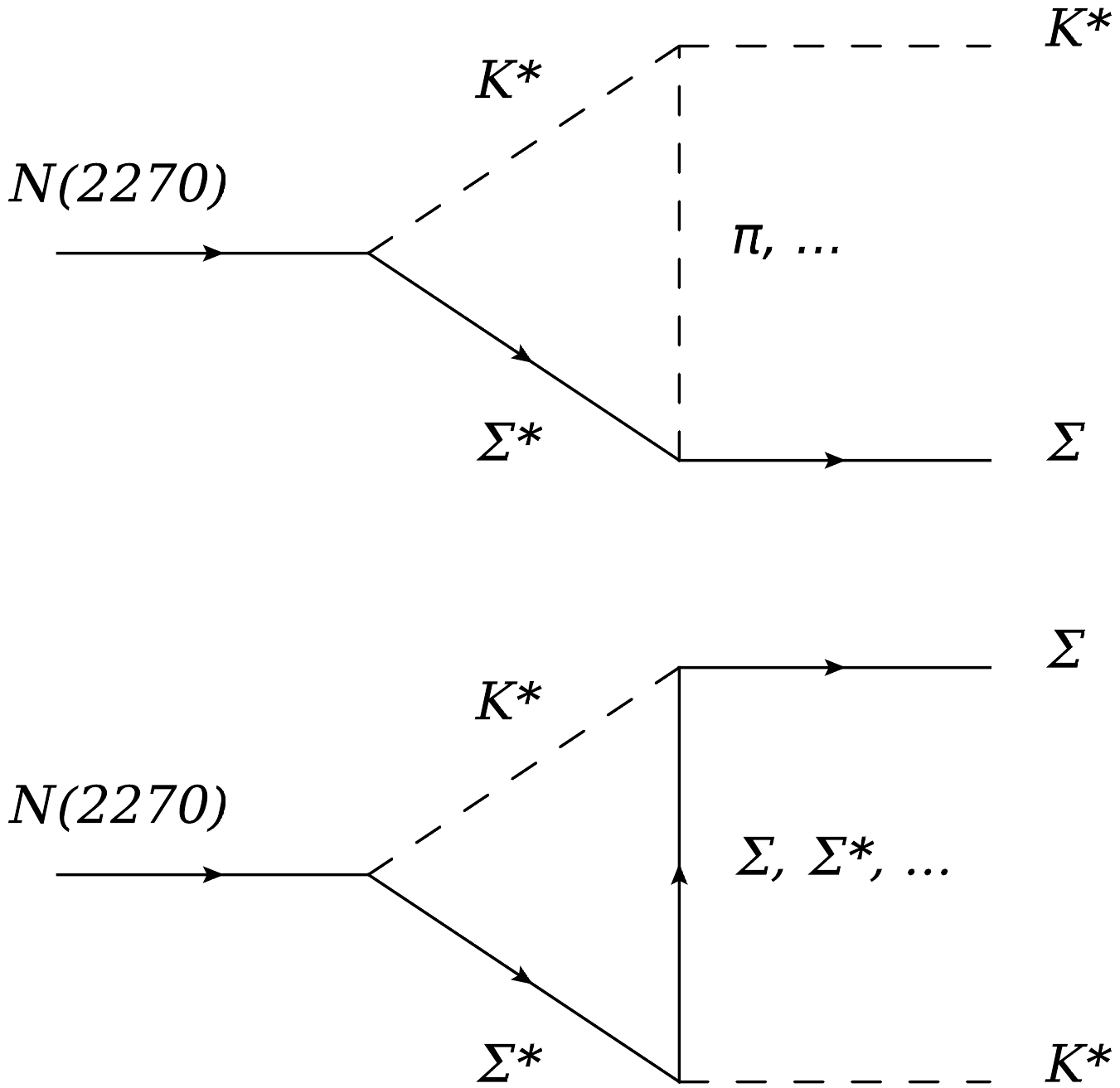}
    \vglue 6pt
    \caption{Hadronic coupling of $N(2270){3/2}^-$ as $K^\ast \Sigma^\ast$ molecule to $K^\ast\Sigma$.}
    \label{2270}
\end{figure}

In the case of $N(2270)3/2^-$, which is assumed as a molecular state of $K^*$ and $\Sigma^*$, the mass is taken to be $M_R=2270$ MeV \cite{Wu:2023}. The hadronic coupling of $K^*\Sigma$ and $N(2270){3/2}^-$ in the hadronic molecular picture is, in principle, dedicated by the loop diagram illustrated in Fig.~\ref{2270}. Here for simplicity, we use the effective Lagrangian given in Eq.~(\ref{hms}) to calculate the $N(2270)\Sigma K^{\ast}$ vertex. The hadronic coupling constant multiplied by the corresponding electromagnetic coupling constant is fixed by fitting the data, as only the product of them is proportional to the reaction amplitudes and relevant in our calculation.

\begin{figure}[tb]
    \centering
    \includegraphics[width=0.3\textwidth]{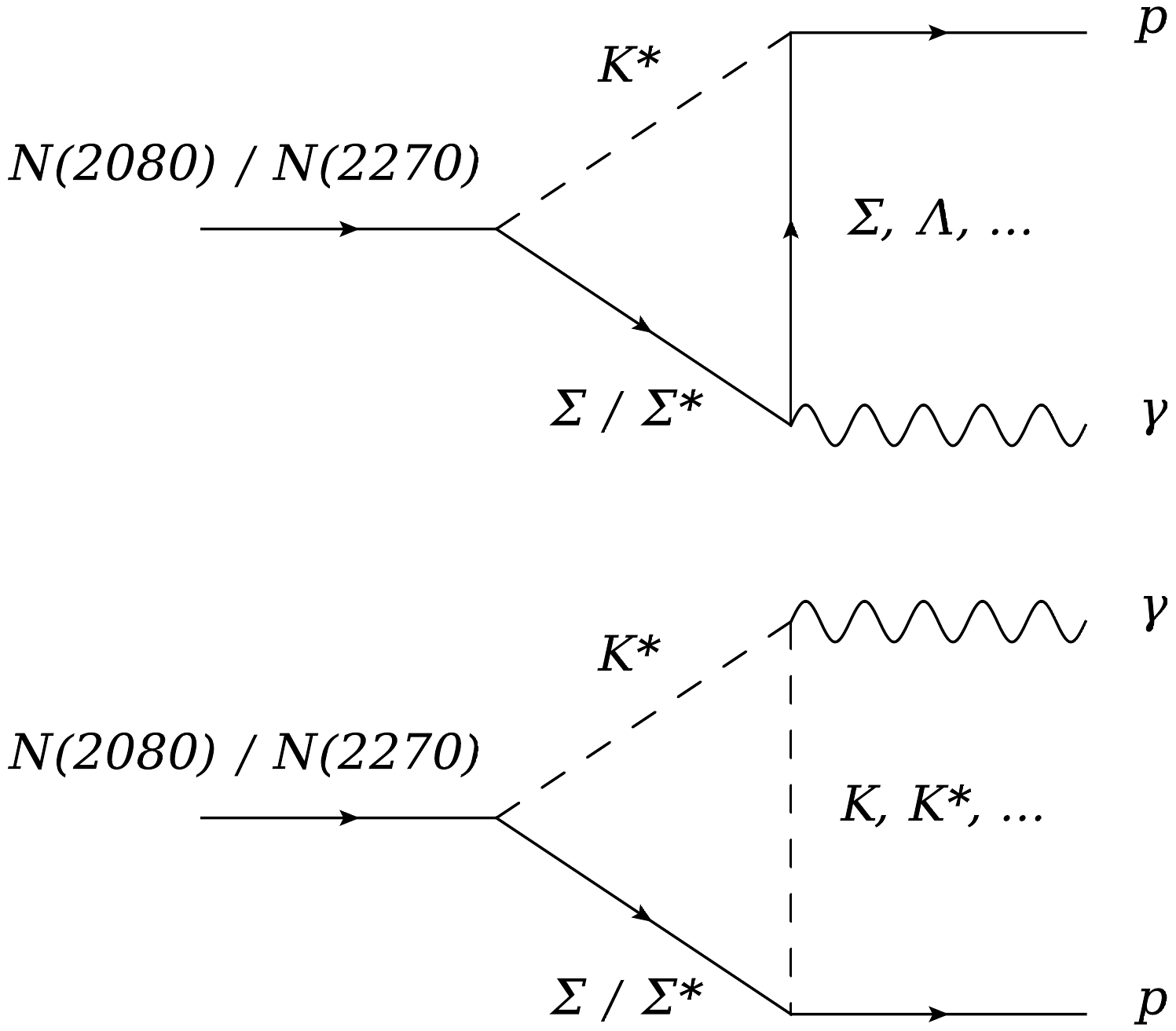}
    \vglue 6pt
    \caption{Electromagnetic coupling of $N(2080){3/2}^-$ as $K^\ast \Sigma$ molecule.}
    \label{fig2}
\end{figure}

The electromagnetic couplings of $N(2080){3/2}^-$ and $N(2270)3/2^-$ in the hadronic molecular picture are, in principle, dedicated by the loop diagram illustrated in Fig.~\ref{fig2}. Here for simplicity, we introduce an effective Lagrangian for $N(2080){3/2}^-$ and $N(2270)3/2^-$ coupling with $N\gamma$:
\begin{eqnarray}
    \mathcal{L}_{\gamma NR} &=& -\, i e\frac{g^{(1)}_{R N\gamma}}{2M_N}\bar{R_\mu}\gamma_\nu  F^{\mu \nu} N \nonumber \\
     && + \, e\frac{g^{(2)}_{R N \gamma}}{\left(2M_N\right)^2} \bar{R_\mu}  F^{\mu \nu}\partial_\nu N + \hc. \label{eq:nnr}
\end{eqnarray}
Then the electromagnetic vertices of $N(2080){3/2}^-$ and $N(2270)3/2^-$ are approximated by calculating the tree-level Feynman diagrams from this Lagrangian, and an additional phase factor ${\rm Exp}[i\phi_R]$ is attached in front of the amplitude resulted from each of the $s$-channel $N(2080){3/2}^-$ and $N(2270)3/2^-$ exchanges to partially mimic the corresponding loop contributions as illustrated in Fig.~\ref{fig2}. Here $\phi_R$ is treated as a fit parameter. In practical calculation, apart from the ratio $g^{(2)}_{R N\gamma}/g^{(1)}_{R N\gamma}$ for both $N(2080){3/2}^-$ and $N(2270)3/2^-$, the coupling $g^{(1)}_{R N\gamma}$ for $N(2080)3/2^-$ and the product $g_{K^\ast\Sigma R} g^{(1)}_{R N\gamma}$ for $N(2270)3/2^-$ are treated as fitting parameters. In Ref.~\cite{Lin:2018kcc}, it is shown that the calculated widths of hadronic molecules depend on the choice of the cutoff parameters. Here we treat the widths of both $N(2080){3/2}^-$ and $N(2270)3/2^-$ as fit parameters.

\subsection{Single spin observables} \label{Sec:asy}

Following Refs.~\cite{Fasano:1992,Sandorfi:2010uv}, the single-polarization observables of photon beam asymmetry ($\Sigma$), target nucleon asymmetry ($T$), and recoil nucleon asymmetry ($P$) are defined as
\begin{align}
\Sigma & \,=\, \dfrac{\dfrac{\,{\rm d}\sigma}{{\rm d}\Omega}(\perp,0,0) \,-\, \dfrac{\,{\rm d}\sigma}{{\rm d}\Omega}(\parallel,0,0)}{\dfrac{\,{\rm d}\sigma}{{\rm d}\Omega}(\perp,0,0) \,+\, \dfrac{\,{\rm d}\sigma}{{\rm d}\Omega}(\parallel,0,0)}, \label{eq:beam}  \\[6pt]
T & \,=\, \dfrac{\dfrac{\,{\rm d}\sigma}{{\rm d}\Omega}(0,+y,0) \,-\, \dfrac{\,{\rm d}\sigma}{{\rm d}\Omega}(0,-y,0)}{\dfrac{\,{\rm d}\sigma}{{\rm d}\Omega}(0,+y,0) \,+\, \dfrac{\,{\rm d}\sigma}{{\rm d}\Omega}(0,-y,0)}, \label{eq:target}  \\[6pt]
P & \,=\, \dfrac{\dfrac{\,{\rm d}\sigma}{{\rm d}\Omega}(0,0,+y) \,-\, \dfrac{\,{\rm d}\sigma}{{\rm d}\Omega}(0,0,-y)}{\dfrac{\,{\rm d}\sigma}{{\rm d}\Omega}(0,0,+y) \,+\, \dfrac{\,{\rm d}\sigma}{{\rm d}\Omega}(0,0,-y)}. \label{eq:recoil}
\end{align}
Here the three arguments of ${{\rm d}\sigma}/{{\rm d}\Omega}$ denote the polarizations of the beam photon, target nucleon, and recoil $\Sigma$ baryon, respectively. The symbols ``$\perp$" and ``$\parallel$" denote that the photon beam is linearly polarized perpendicular and parallel to the reaction plane, respectively. The symbols ``$+y$" and ``$-y$" denote that the target nucleon or recoil $\Sigma$ baryon is polarized along the directions of ${\bm k}\times{\bm q}$ and $-\left({\bm k}\times{\bm q}\right)$, respectively, with ${\bm k}$ and ${\bm q}$ being the three-momentum of incoming photon and outgoing $K^\ast$. The symbol ``0" denotes that the corresponding argument is unpolarized.

\section{Results and Discussion}\label{Sec:results}

As has been mentioned in Sec.~\ref{Sec:intro}, in literature the most recent and comprehensive investigation of the $\gamma p \to K^{\ast +} \Sigma^0$ and $\gamma p \to K^{\ast 0} \Sigma^+$ reactions is the one from Ref.~\cite{Wang:2018vlv}, where all the available differential and total cross-section data for $K^\ast \Sigma$ photoproduction off proton have been analyzed in an effective Lagrangian approach with the $\Delta(1905){5/2}^+$, a four-star resonance advocated in the most recent PDG review \cite{PDG2022}, being considered. It was found in Ref.~\cite{Wang:2018vlv} that the cross sections of $\gamma p \to K^{\ast +} \Sigma^0$ are dominated by $s$-channel $\Delta(1905){5/2}^+$ exchange at low energies and $t$-channel $K^\ast$ exchange at high energies, while for the $\gamma p \to K^{\ast 0} \Sigma^+$ reaction, the angular dependences are dominated by $t$-channel $K$ exchange at forward angles and $u$-channel $\Sigma^\ast$ exchange at backward angles.

In the present work, we re-analyze the $\gamma p \to K^{\ast +} \Sigma^0$ and $\gamma p \to K^{\ast 0} \Sigma^+$ reactions by considering the contributions from the $N(2080){3/2}^-$ molecule which was proposed to be the strange partner of $P_c(4457)$ \cite{He:2017aps,Lin:2018kcc}, and the $N(2270)3/2^-$ molecule which is an analogue in the light quark sector of the $\bar{D}^\ast \Sigma^\ast_c$ bound states that are predicted as members in the same heavy-quark spin symmetry multiplet as the $P_c$ states \cite{Liu:2019}. The purpose is to check whether the available differential and total cross-section data of $K^\ast \Sigma$ photoproduction off proton are compatible with the molecular scenario of $N(2080){3/2}^-$ as a $K^\ast \Sigma$ shallowly bound state and $N(2270)3/2^-$ as a $K^\ast\Sigma^\ast$ shallowly bound state. We consider all the available data for $K^\ast \Sigma$ photoproduction from the $K^\ast \Sigma$ threshold ($\sim 2086$ MeV) up to the center-of-mass energy $W=2.8$ GeV. Note that $K^\ast \Sigma$ can couple to $N(2080){3/2}^-$ and $N(2270)3/2^-$ in $S$ wave, while it couples to $\Delta(1905){5/2}^+$ in $P$ wave or even higher odd partial waves. In this sense, the $N(2080){3/2}^-$ and $N(2270)3/2^-$ might have stronger effects than $\Delta(1905){5/2}^+$ in $K^\ast \Sigma$ photoproduction reactions.

\begin{table}[tb]
\caption{\label{tab:new_set} Fitted values of free model parameters.}
\begin{tabular*}{0.8\columnwidth}{@{\extracolsep\fill}lr}
\hline\hline
        $g^{(1)}_{\Sigma^{*0}\Sigma^0\gamma}$ & $7.06 \pm 2.55$   \\
        $g^{(2)}_{\Sigma^{*0}\Sigma^0\gamma}$ & $-38.83 \pm 11.15$   \\
        $g^{(1)}_{\Delta \Sigma K^\ast}$ & $-0.42 \pm 0.14$   \\
        $g^{(1)}_{N(2080)N\gamma}$ & $ -0.12 \pm 0.04 $   \\
        $g^{(2)}_{N(2080)N\gamma}/g^{(1)}_{N(2080)N\gamma}$ & $ -1.60 \pm 0.19 $   \\
        $g^{(1)}_{N(2270)N\gamma} g_{K^\ast\Sigma N(2270)}$ & $0.28 \pm 0.06$   \\
        $g^{(2)}_{N(2270)N\gamma}/g^{(1)}_{N(2270)N\gamma}$ & $ -0.51 \pm 0.12 $   \\
        $\phi_{N(2080)}$ & $2.83 \pm 0.26$   \\
        $\phi_{N(2270)}$ & $1.55 \pm 0.13$   \\
        $\Gamma_{N(2080)}$ [MeV] & $70.1 \pm 9.7$   \\
        $\Gamma_{N(2270)}$ [MeV] & $361.3 \pm 8.9$   \\
        $\Lambda_{R}$ [MeV]  & $1607\pm 118$   \\
        $\Lambda_s$ [MeV] & $1862\pm 31$   \\
        $\Lambda_t$ [MeV] & $1064\pm 26$   \\
        $\Lambda_u$ [MeV] & $715\pm 35$   \\
\hline\hline
\end{tabular*}
\end{table}

\begin{figure*}[htb]
    \centering
    \includegraphics[width=0.9\textwidth]{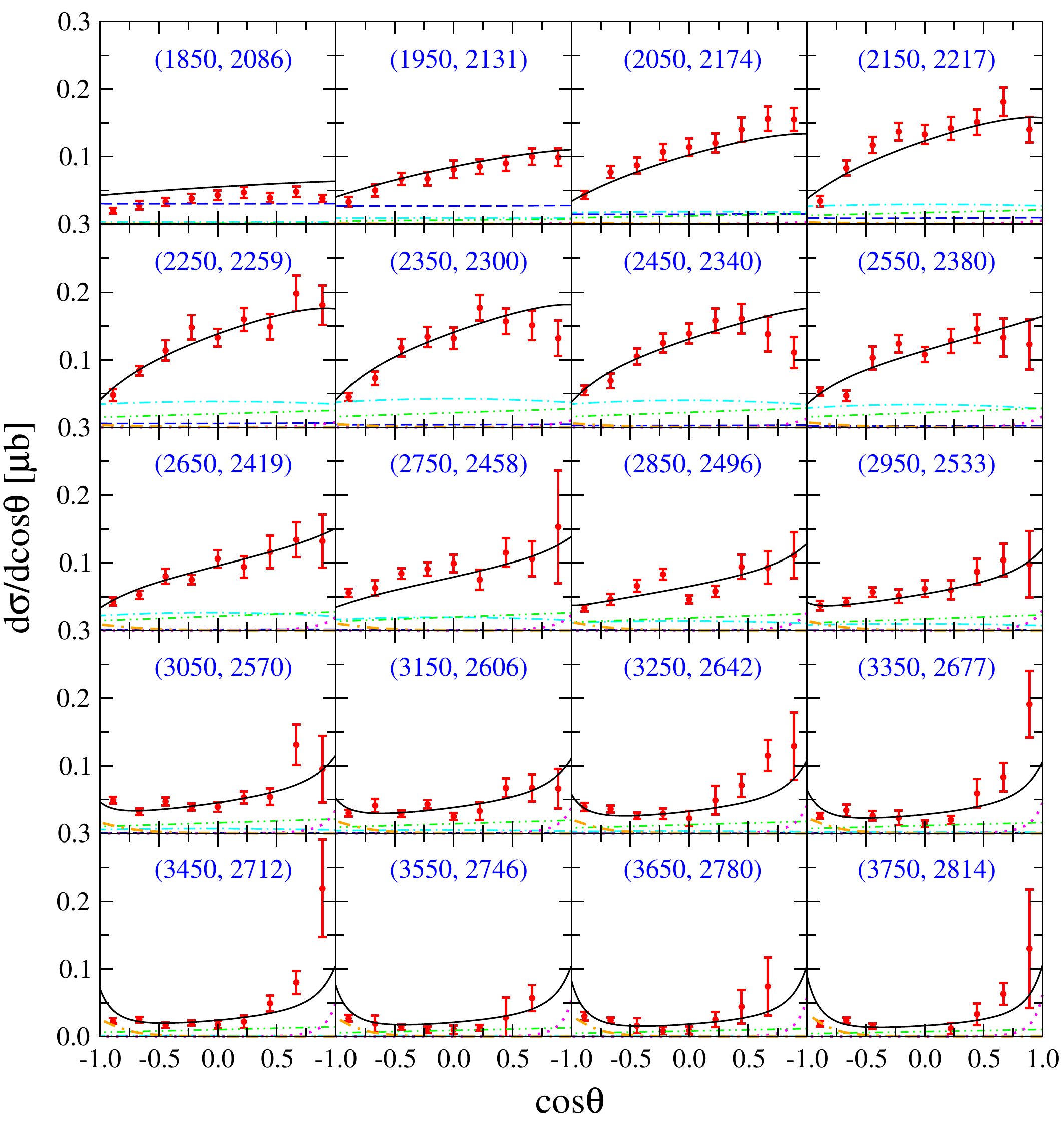}
    \caption{Differential cross sections for $\gamma p \rightarrow K^{\ast +} \Sigma^0$ as a function of $\cos\theta$. The numbers in parentheses denote the photon laboratory incident energy (left number) and the total center-of-mass energy of the system (right number). The blue dashed lines, cyan dash-dotted lines, and green double-dotted lines represent the individual contributions from the $s$-channel $N(2080){3/2}^-$, $N(2270){3/2}^-$, and $N$ exchanges, respectively. The magenta dotted lines and orange double-dash-dotted lines represent the individual contributions from the $t$-channel $K^*$ exchange and $u$-channel $\Sigma^\ast$ exchange, respectively. The scattered symbols denote the CLAS data in Ref.~\cite{Wei:2013}. }
    \label{fig:diff-kp}
\end{figure*}

\begin{figure*}[htb]
    \centering
    \includegraphics[angle=270, width=0.9\textwidth]{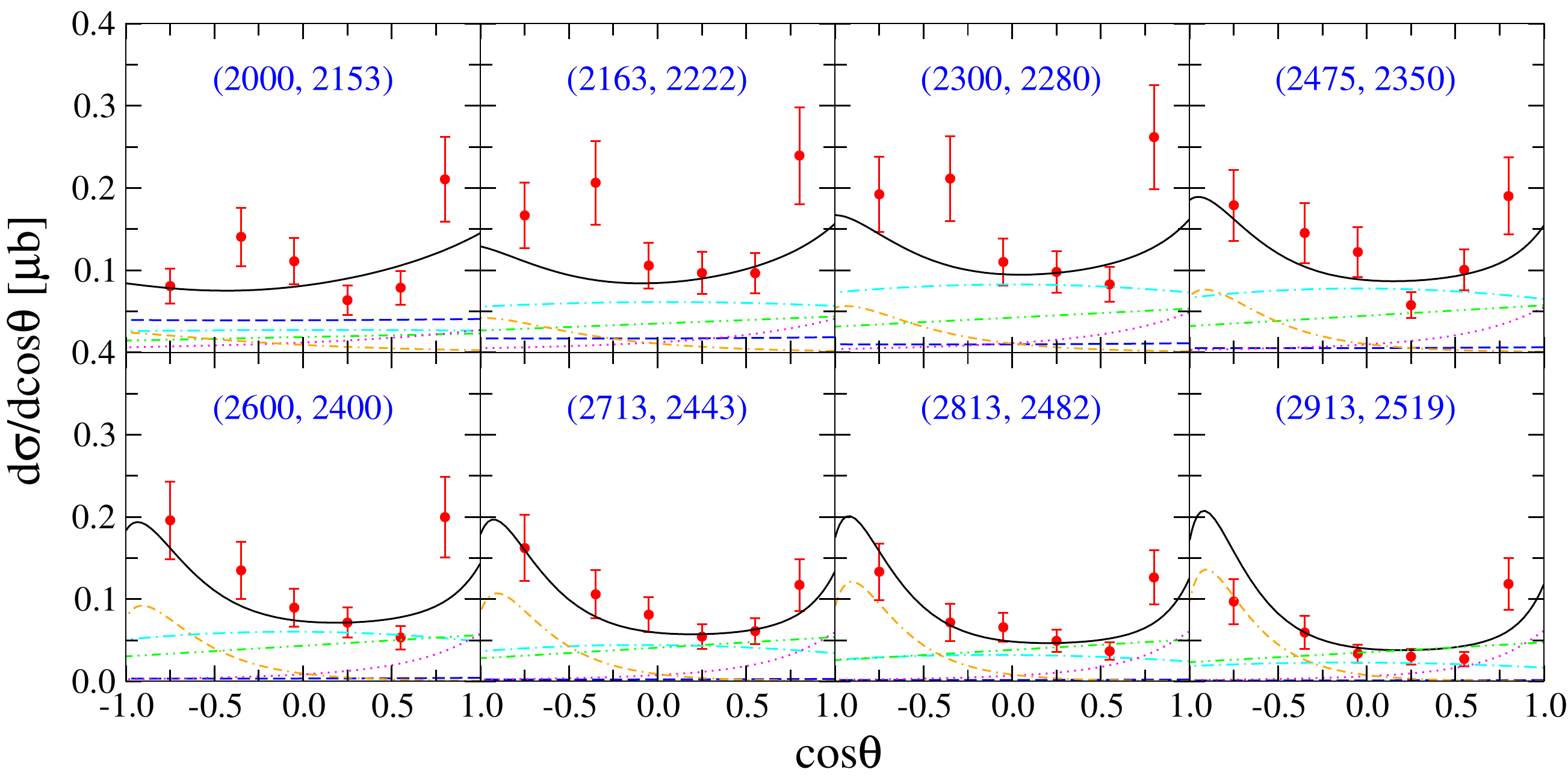}
    \caption{Differential cross sections for $\gamma p \rightarrow K^{\ast 0} \Sigma^+$ as a function of $\cos\theta$. Notations are the same as in Fig.~\ref{fig:diff-kp} except that now the magenta dotted lines represent the individual contributions from the $t$-channel $K$ exchange and the scattered symbols denote the CLAS data in Ref.~\cite{Hleiqawi:2007ad}. }
    \label{fig:diff-k0}
\end{figure*}

\begin{figure}[tb]
    \centering
    \subfigure[~$\gamma p \rightarrow K^{\ast +}\Sigma^0$]{
    \includegraphics[angle=270, width=0.95\columnwidth]{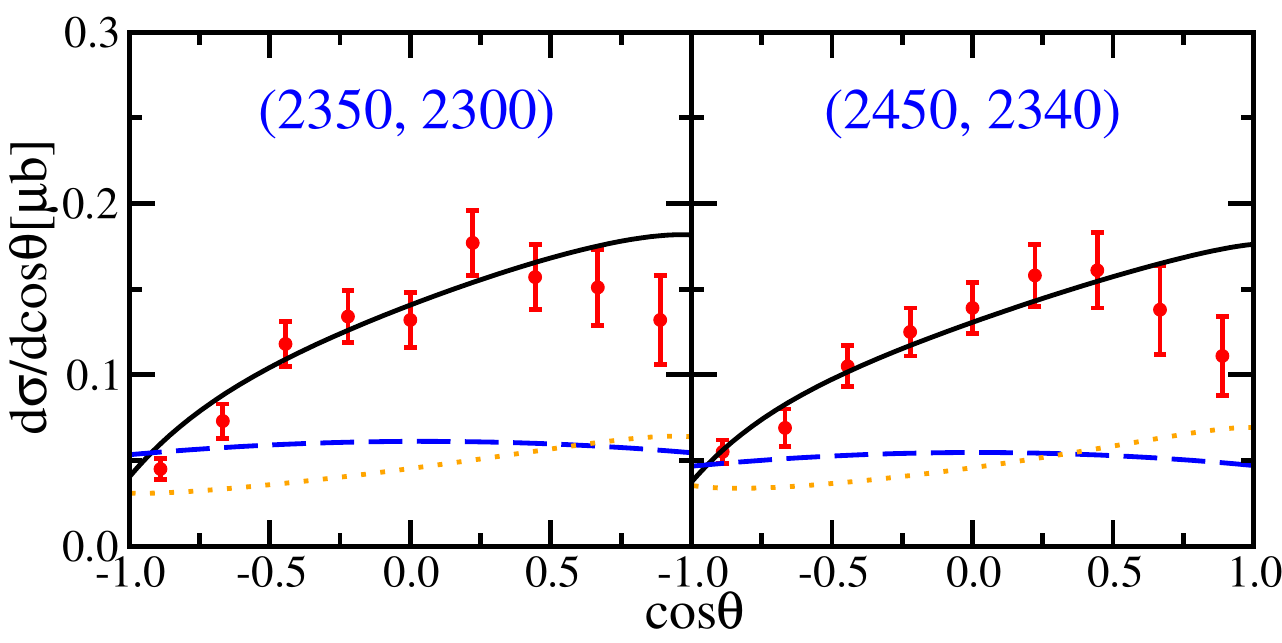}}  \\[1pt]
    \subfigure[~$\gamma p \rightarrow K^{\ast 0}\Sigma^+$]{
    \includegraphics[angle=270, width=0.95\columnwidth]{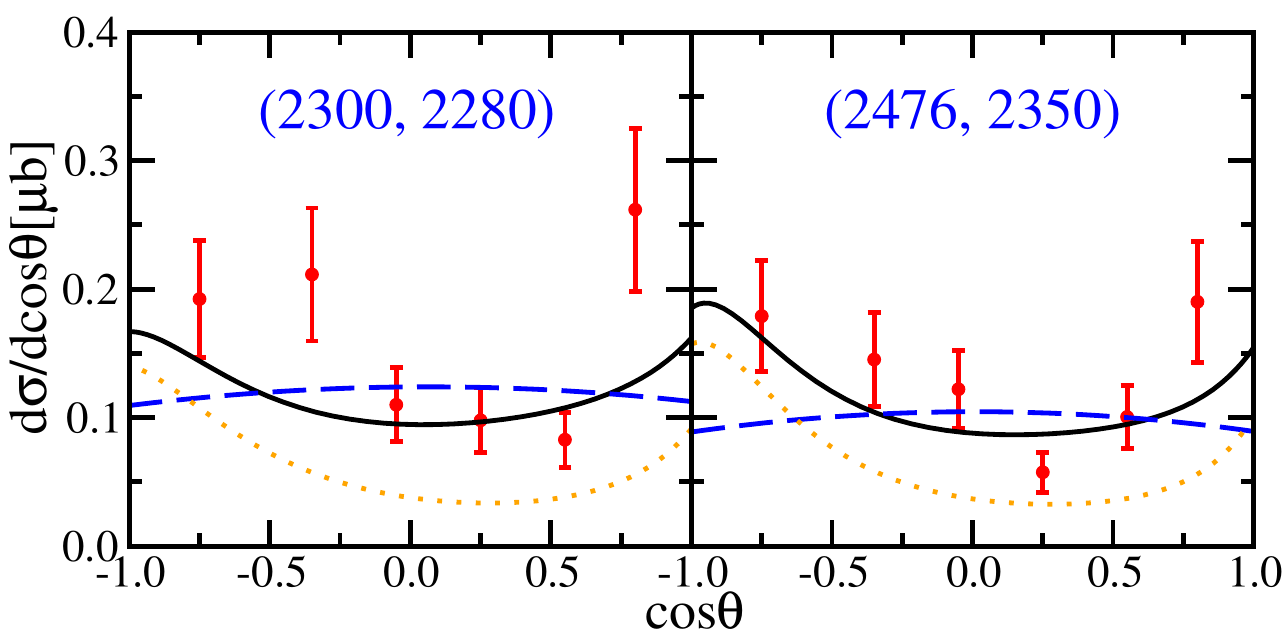}}
    \caption{Differential cross sections as functions of $\cos\theta$ for (a) $\gamma p \rightarrow K^{\ast +} \Sigma^0$ and (b) $\gamma p \rightarrow K^{\ast 0} \Sigma^+$. The numbers in parentheses denote the photon laboratory incident energy (left number) and the total center-of-mass energy of the system (right number). The black solid lines represent the full results. The blue dashed lines represent the coherent sum of contributions from the $s$-channel $N(2080){3/2}^-$ and $N(2270){3/2}^-$ exchanges. The orange dotted lines represent the results calculated by switching off the contributions of the $s$-channel $N(2080){3/2}^-$ and $N(2270){3/2}^-$ exchanges.}
    \label{fig:int}
\end{figure}

\begin{figure*}[tb]
    \centering
    {\vglue 0.15cm}
    \subfigure{
    \includegraphics[angle=270, width=0.9\columnwidth]{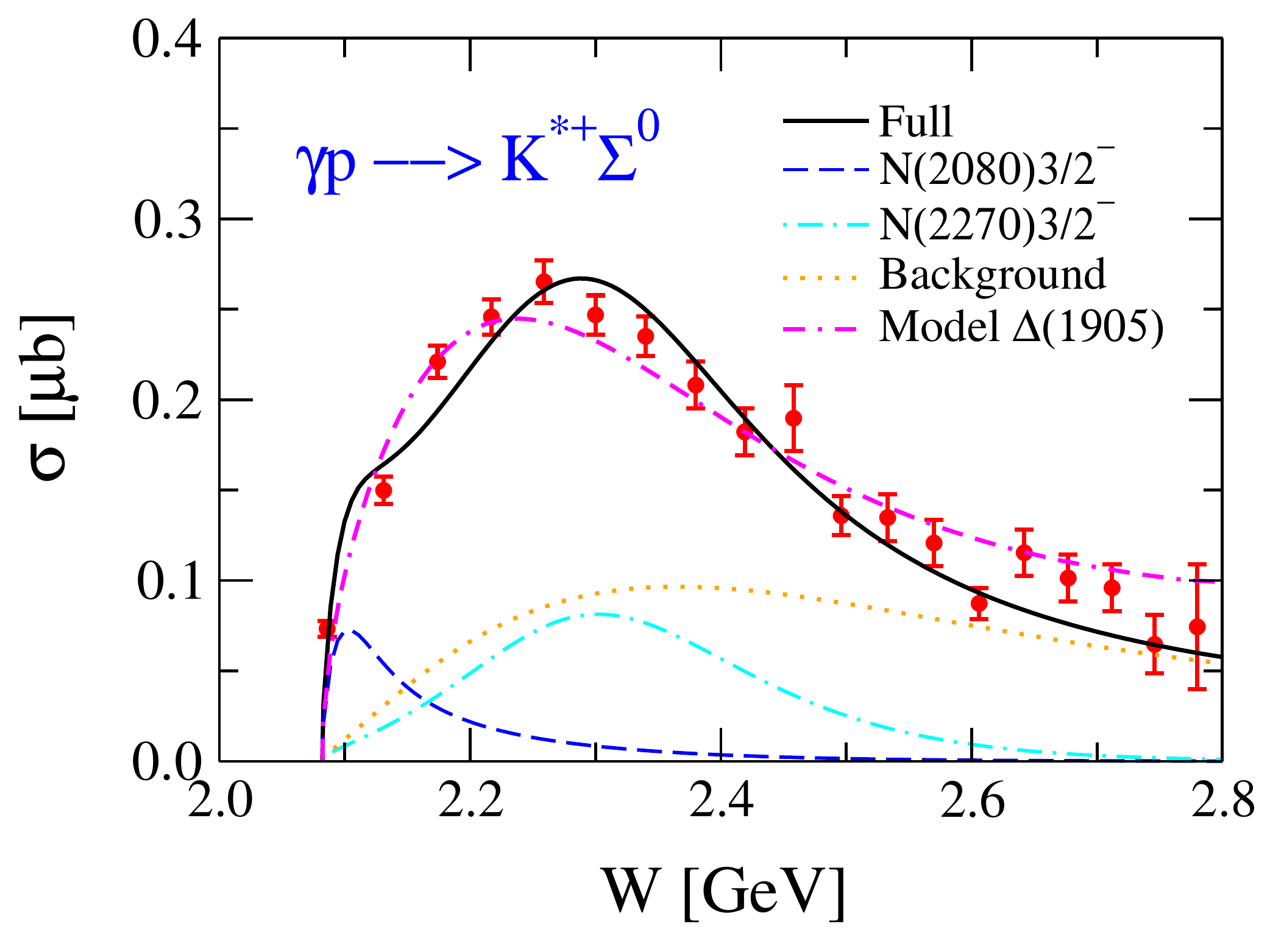}} {\hglue 0.4cm}
    \subfigure{
    \includegraphics[angle=270, width=0.9\columnwidth]{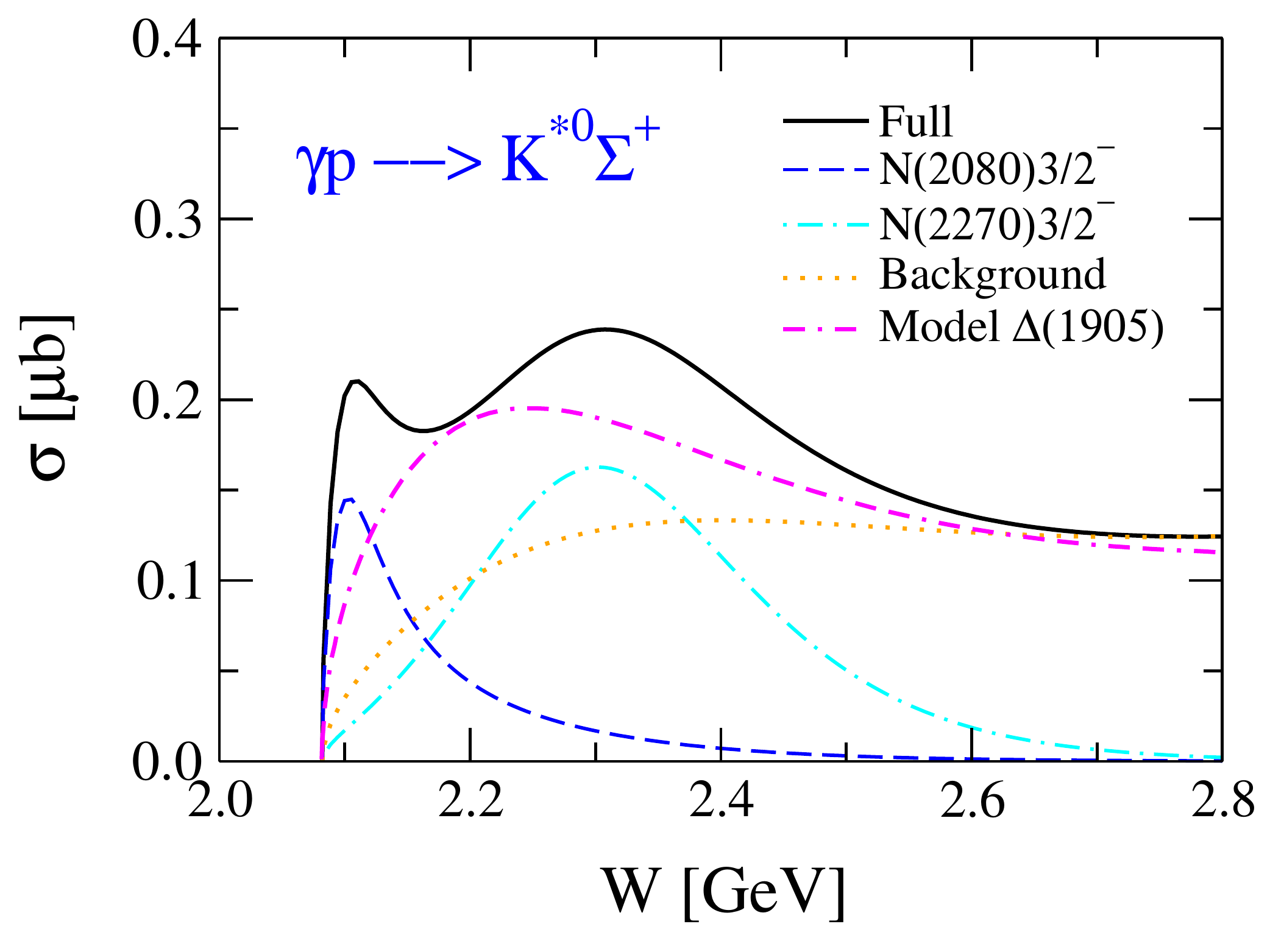}}
    \caption{Total cross sections for $\gamma p \rightarrow K^{\ast +}\Sigma^0$ (left) and $\gamma p \rightarrow K^{\ast 0}\Sigma^+$ (right). The black solid lines represent the full results. The blue dashed lines and cyan dash-dotted lines represent the individual contributions from the $s$-channel $N(2080){3/2}^-$ and $N(2270){3/2}^-$ exchanges, respectively. The orange dotted lines represent the results calculated by switching off the contributions from the $N(2080){3/2}^-$ and $N(2270){3/2}^-$ exchanges. The magenta double-dash-dotted lines represent the full results of Ref.~\cite{Wang:2018vlv}. The scattered symbols are data from CLAS Collaboration \cite{Wei:2013}.}
    \label{fig:total}
\end{figure*}

\begin{figure*}[htb]
    \centering
    \includegraphics[width=0.9\textwidth]{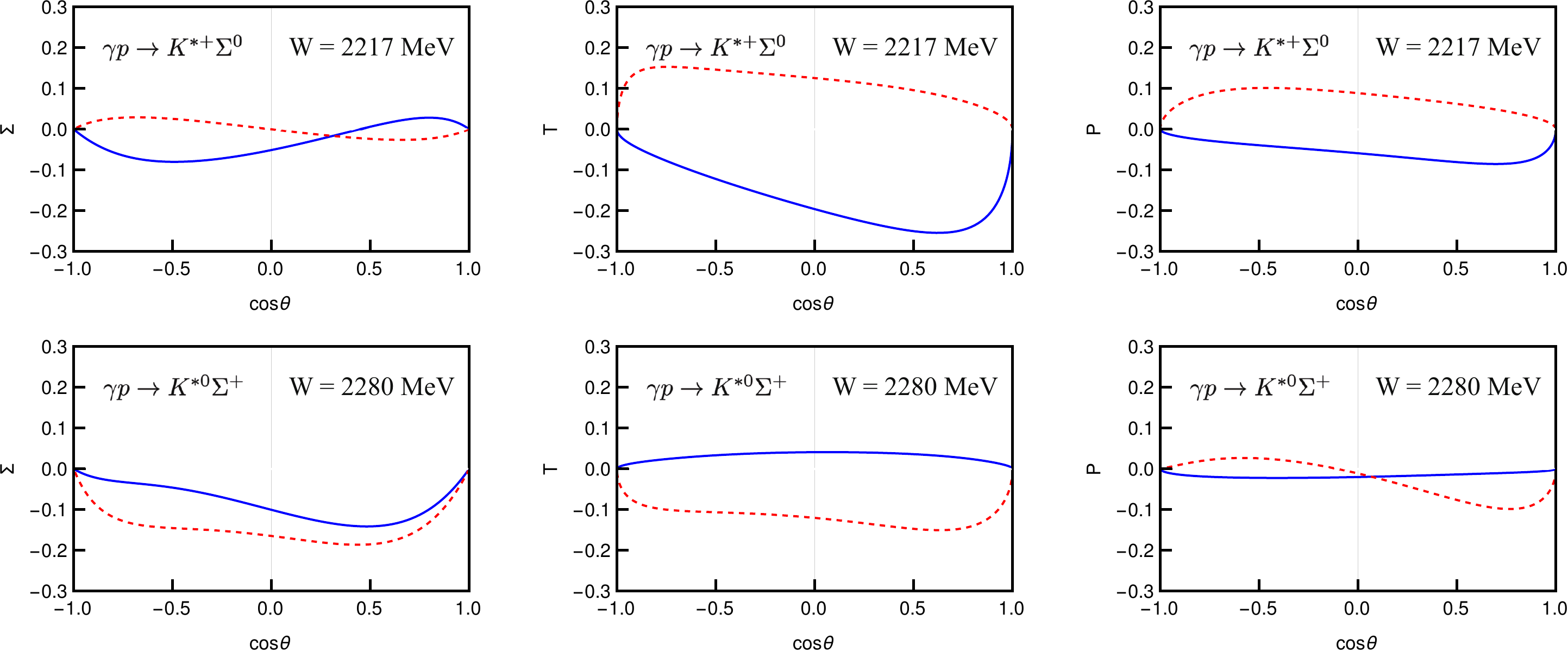}
    \caption{Single spin asymmetries $\Sigma$ (left), $T$ (middle), and $P$ (right) predicted at $W=2217$ MeV for $\gamma p\rightarrow K^{\ast +}\Sigma^0$ (the upper row) and $W=2280$ MeV for $\gamma p \rightarrow K^{\ast 0}\Sigma^+$ (the lower row). The blue solid lines represent the results from the present work, and the red dashed lines denote the results from Ref.~\cite{Wang:2018vlv}. }
    \label{fig:asy}
\end{figure*}

In Ref.~\cite{Wang:2018vlv}, different cutoff parameters were employed for $\Lambda$, $\Sigma$, $\Sigma^\ast$, $K$, $\kappa$, and $K^\ast$ exchanges. In the present work, in order to reduce the number of fit parameters, we use the same cutoff parameter $\Lambda_u$ for the $u$-channel $\Lambda$, $\Sigma$, and $\Sigma^\ast$ exchanges, and the same cutoff parameter $\Lambda_t$ for the $t$-channel $K$, $\kappa$, and $K^\ast$ exchanges. The effective Lagrangian for the electromagnetic coupling of  $u$-channel $\Sigma^\ast$ exchange reads
\begin{eqnarray}
{\cal L}_{\Sigma^\ast \Sigma \gamma} &=& ie\frac{g^{(1)}_{\Sigma^\ast \Sigma \gamma}}{2M_N}\bar{\Sigma}\gamma_\nu \gamma_5 F^{\mu \nu} \Sigma^\ast_\mu \nonumber \\
&& -\,e\frac{g^{(2)}_{\Sigma^\ast \Sigma \gamma}}{\left(2M_N\right)^2} \left(\partial_\nu\bar{\Sigma}\right) \gamma_5 F^{\mu \nu}{\Sigma}^\ast_\mu + \hc.
\label{Sigmastar}
\end{eqnarray}
For $\Sigma^{\ast +}$ exchange, the PDG review \cite{PDG2022} gives $\Gamma_{\Sigma^{\ast +}\rightarrow \Sigma^+ \gamma} = 0.252$ MeV, which results in one constraint for the two couplings $g^{(1)}_{\Sigma^\ast \Sigma \gamma}$ and $g^{(2)}_{\Sigma^\ast \Sigma \gamma}$, and only lets one of these two couplings as free parameter. In Ref.~\cite{Wang:2018vlv}, the $g^{(2)}_{\Sigma^{\ast +} \Sigma^+ \gamma}/g^{(1)}_{\Sigma^{\ast +} \Sigma^+ \gamma}$ was set as a fit parameter and then the $g^{(1)}_{\Sigma^{\ast +} \Sigma^+ \gamma}$ was fixed by use of the value of $\Gamma_{\Sigma^{\ast +}\rightarrow\Sigma^+\gamma}$ \cite{PDG2022}. In the present work, we simply set $g^{(1)}_{\Sigma^{\ast +} \Sigma^+ \gamma}=0$ and fit the $g^{(2)}_{\Sigma^{\ast +} \Sigma^+ \gamma}$ by the data of $\Gamma_{\Sigma^{\ast +}\rightarrow\Sigma^+\gamma}$, as the fitting process tells us that the contributions from the $g^{(1)}_{\Sigma^{\ast +} \Sigma^+ \gamma}$ term is negligible. The resulted value of $g^{(2)}_{\Sigma^{\ast +} \Sigma^+ \gamma}$ reads $g^{(2)}_{\Sigma^{\ast +} \Sigma^+ \gamma} = 76.793$.

The free model parameters and their fitted values are listed in Table~\ref{tab:new_set}. There, $g^{(1)}_{\Delta\Sigma K^\ast}$ is the hadronic coupling constant for $\Delta$ pole diagram. $g^{(1)}_{R N \gamma}$ and $g^{(2)}_{R N \gamma}$ are the electromagnetic coupling constants for pole diagram of molecular state $R$ exchange, where $R \equiv N(2080){3/2}^-$ or $N(2270){3/2}^-$. $\phi_{N(2080)}$ and $\phi_{N(2270)}$ are the parameters in phase factors ${\rm Exp}[i\phi_{N(2080)}]$ and ${\rm Exp}[i\phi_{N(2270)}]$ attached in front of the amplitudes resulted from $N(2080)3/2^-$ and $N(2270){3/2}^-$ pole diagrams, respectively. $\Gamma_{N(2080)}$ and $\Gamma_{N(2270)}$ are the widths of the $N(2080)3/2^-$ and $N(2270){3/2}^-$ states, respectively. The uncertainties of the fitted values of the parameters are estimates arising from the uncertainties (error bars) associated with the fitted data points. The obtained chi-squared ($\chi^2$) per data point is $1.648$, indicating a good fitting quality of the theoretical results. Note that our fitted decay width of $N(2080)3/2^-$ is $70.1$ MeV, smaller than the value $141.1$ MeV obtained by calculating the partial decay widths of various decay channels in an effective Lagrangian approach in Ref.~\cite{Lin:2018kcc}, although the same mass of $N(2080)3/2^-$ is adopted in both of these two works. We mention that the width calculated in Ref.~\cite{Lin:2018kcc} changes from $50$ to $350$ MeV when the cutoffs in form factors vary in a reasonable range, as shown in Fig.~4 of Ref.~\cite{Lin:2018kcc}.

The results of differential cross sections for $\gamma p \rightarrow K^{\ast +}\Sigma^0$ and $\gamma p \rightarrow K^{\ast 0}\Sigma^+$ corresponding to the parameters listed in Table~\ref{tab:new_set} are shown in Fig.~\ref{fig:diff-kp} and Fig.~\ref{fig:diff-k0}, respectively. There, the numbers in parentheses denote the photon laboratory incident energy (left number) and the total center-of-mass energy of the system (right number). The black solid lines represent the results from the full amplitudes. The blue dashed lines, cyan dash-dotted lines, and green double-dotted lines represent the individual contributions from the $s$-channel $N(2080){3/2}^-$, $N(2270)3/2^-$, and $N$ exchanges, respectively. The orange double-dash-dotted lines represent the individual contributions from the $u$-channel $\Sigma^\ast$ exchange. The magenta dotted lines denote the individual contributions from the $t$-channel $K^\ast$ exchange in Fig.~\ref{fig:diff-kp} and $K$ exchange in Fig.~\ref{fig:diff-k0}. The contributions from other single terms are too small to be clearly seen with the scale used, and thus they are not plotted. One sees from Fig.~\ref{fig:diff-kp} and Fig.~\ref{fig:diff-k0} that our overall description of the CLAS angular distribution data for $\gamma p \rightarrow K^{\ast +}\Sigma^0$ and $\gamma p \rightarrow K^{\ast 0}\Sigma^+$ in  the full energy region is fairly satisfactory. Compared with the results from Ref.~\cite{Wang:2018vlv}, for the $\gamma p \rightarrow K^{\ast +}\Sigma^0$ reaction, the fitting quality is similar, while for the $\gamma p \rightarrow K^{\ast 0}\Sigma^+$ reaction, the fitting quality is now improved significantly.

For $\gamma p \rightarrow K^{\ast +}\Sigma^0$, Fig.~\ref{fig:diff-kp} shows that the $s$-channel $N(2080){3/2}^-$ exchange provides dominant contributions to the differential cross sections in the near-threshold energy region, and the $N(2270){3/2}^-$ exchange provides significant contributions in a much wider energy range up to $W\sim 2.5$ GeV. In higher energy region, significant contributions at forward angles raise form the $t$-channel $K^\ast$ exchanges and at backward angles from the $u$-channel $\Sigma^\ast$ exchange. This is quite different from the reaction mechanism reported in Ref.~\cite{Wang:2018vlv}, where it was found that the $s$-channel $\Delta(1905){5/2}^+$ exchange dominates the angular distributions in the near-threshold energy region, and the $s$-channel $\Delta$ exchange and $t$-channel $K^\ast$ exchange provide considerable contributions also in the low and high energy region, respectively. The contributions from the $s$-channel $\Delta$ exchange in the present work are much smaller than those in Ref.~\cite{Wang:2018vlv}. This can be understood if one notices that the fitted value of the magnitude of the coupling constant $g^{(1)}_{\Delta \Sigma K^\ast}$ is $0.42$ in the present work, much smaller than the value $8.84$ obtained in Ref.~\cite{Wang:2018vlv}. The contributions from the $s$-channel $N$ exchange in the present work are rather significant, while they are negligible in Ref.~\cite{Wang:2018vlv}. The contributions from the $t$-channel $K^\ast$ exchange provides considerable contributions in Ref.~\cite{Wang:2018vlv}, while they are negligible in the present work. Both these properties for $N$ and $K^\ast$ exchanges can be understood by different values of the fitted cutoff parameters in the present work and Ref.~\cite{Wang:2018vlv}.

For $\gamma p \rightarrow K^{\ast 0}\Sigma^+$, Fig.~\ref{fig:diff-k0} shows that the dominant contributions to the differential cross sections in the low energy region are coming from the $s$-channel $N(2080){3/2}^-$ and $N(2270)3/2^-$ exchanges. The $s$-channel $N$ exchange, $u$-channel $\Sigma^\ast$ exchange, and $t$-channel $K$ exchange also provide significant contributions, especially in the high energy region. In particular, the $s$-channel $N(2080){3/2}^-$ exchange is seen to provide rather important contributions to the differential cross sections at $W=2153$ MeV, while its contributions are relative small at the other energy points. The $s$-channel $N(2270){3/2}^-$ exchange provides the most important contributions to the differential cross sections in the energy range $W\sim 2.2-2.4$ GeV. With the energy increasing, the $u$-channel $\Sigma^\ast$ exchange and the $t$-channel $K$ exchange provide more and more important contributions to the cross sections at backward angles and forward angles, respectively. In Ref.~\cite{Wang:2018vlv}, it was reported that the angular distributions are dominated by the $t$-channel $K$ exchange at forward angles and the $u$-channel $\Sigma^\ast$ exchange at backward angles, the $s$-channel $\Delta(1905){5/2}^+$ exchange makes considerable contributions at low energies, and the $s$-channel $\Delta$ exchange gives small but non-negligible contributions near threshold. The $\Delta(1905){5/2}^+$ contributes in a much wider energy range in Ref.~\cite{Wang:2018vlv} as this resonance has a relatively large width, $\Gamma_{\Delta(1905){5/2}^+}\approx 330$ MeV. In the present work, the contributions from $N(2080){3/2}^-$ to the differential cross sections are significantly dominant only at the lowest energy $W=2153$ MeV since the value of the width of $N(2080){3/2}^-$ is fitted to be narrow, $\Gamma_{N(2080)} \approx 70.1$ MeV, as listed in Table~\ref{tab:new_set}. The differences of the contributions from other exchange diagrams in the present work and in Ref.~\cite{Wang:2018vlv} can be understood from the differences of fitted values of the corresponding cutoff parameters and coupling constants. Note that the contributions from $N(2080){3/2}^-$, $N(2270)3/2^-$, and $N$ exchanges to the differential cross sections of $\gamma p \rightarrow K^{\ast 0}\Sigma^+$ in the present work are much bigger than those in Ref.~\cite{Wang:2018vlv}. As a consequence, the theoretical differential cross sections from the present work agree much well with the data than the results of Ref.~\cite{Wang:2018vlv}.

Figure~\ref{fig:int} is plotted to illustrate the interference effects of the resonance and background contributions. The upper row is for the $\gamma p \rightarrow K^{\ast +}\Sigma^0$ process while the lower row is for the $\gamma p \rightarrow K^{\ast 0}\Sigma^+$ process. The black solid lines represent the results from the full amplitudes. The blue dashed lines represent the coherent sum of the contributions from the $s$-channel $N(2080){3/2}^-$ and $N(2270){3/2}^-$ exchanges. The orange dotted lines denote the results calculated by switching off the contributions from the $s$-channel $N(2080){3/2}^-$ and $N(2270){3/2}^-$ exchanges. One sees that in both reactions, the algebraic sum of the contributions from the $N(2080){3/2}^-$ and $N(2270){3/2}^-$ exchanges (blue dashed lines) and the contributions from all the other terms (orange dotted lines) does not match the full results (black solid lines). Compared with Figs.~\ref{fig:diff-kp} and \ref{fig:diff-k0}, one also notices that the algebraic sum of the individual contributions from $N(2080){3/2}^-$ and $N(2270){3/2}^-$ exchanges does not match the coherent sum of them, either. This can be understood if one observes that in the present work, an additional phase factor ${\rm Exp}[i\phi_R]$ is attached in front of the amplitude resulted from each of the $s$-channel $N(2080){3/2}^-$ and $N(2270)3/2^-$ exchanges to partially mimic the corresponding loop contributions as illustrated in Fig.~\ref{fig2}. Therefore, the interference effects of them in the present work would be much more important than in the traditional calculation where only tree level diagrams are calculated and no phase factors are considered.

Figure~\ref{fig:total} shows our predicted total cross sections for $\gamma p \rightarrow K^{\ast +}\Sigma^0$ (left graph) and $\gamma p \rightarrow K^{\ast 0}\Sigma^+$ (right graph) obtained via an integration of the corresponding differential cross sections as shown in Fig.~\ref{fig:diff-kp} and Fig.~\ref{fig:diff-k0}. In this figure, the black solid lines represent the full results. The blue dashed lines and cyan dash-dotted lines represent the individual contributions from the $s$-channel $N(2080){3/2}^-$ exchange and $N(2270){3/2}^-$ exchange, respectively. The orange dotted lines represent the results calculated by switching off the contributions from the $N(2080){3/2}^-$ and $N(2270){3/2}^-$ exchanges. The total cross sections from Ref.~\cite{Wang:2018vlv} are also plotted (magenta double-dash-dotted lines) for comparison. One sees from Fig.~\ref{fig:total} that our predicted total cross sections for $\gamma p \rightarrow K^{\ast +}\Sigma^0$ are in good agreement with the data, and for both $\gamma p \rightarrow K^{\ast +}\Sigma^0$ and $\gamma p \rightarrow K^{\ast 0}\Sigma^+$ reactions, the $s$-channel $N(2080){3/2}^-$ and $N(2270){3/2}^-$ exchanges provide rather important contributions. Compared with Ref.~\cite{Wang:2018vlv}, for $\gamma p \rightarrow K^{\ast +}\Sigma^0$ the total cross sections in these two works are similar, both in agreement with the data, while for $\gamma p \rightarrow K^{\ast 0}\Sigma^+$ the total cross sections in the present work are more structured than those in Ref.~\cite{Wang:2018vlv}. Moreover, near the threshold energy region, the total cross sections for $\gamma p \rightarrow K^{\ast 0}\Sigma^+$ in the present work are much bigger than those for $\gamma p \rightarrow K^{\ast +}\Sigma^0$. While in Ref.~\cite{Wang:2018vlv}, opposite pattern is observed. Unfortunately we don't have data for the total cross sections of $\gamma p \rightarrow K^{\ast 0}\Sigma^+$. But note that the differential cross sections for $\gamma p \rightarrow K^{\ast 0}\Sigma^+$ are described much better in the present work (c.f. Fig.~\ref{fig:diff-k0}) than in Ref.~\cite{Wang:2018vlv}. In this sense, the total cross sections predicted in the present work might be more reliable than those in Ref.~\cite{Wang:2018vlv}. Nevertheless, the available differential cross-section data for $\gamma p\rightarrow K^{\ast 0}\Sigma^+$ have large uncertainties, which might be part of the reason that the total cross sections for this reaction from the present work and Ref.~\cite{Wang:2018vlv} are different. Future high-precision data on differential cross sections and data on total cross sections for $\gamma p\rightarrow K^{\ast 0}\Sigma^+$ are called on to give further insights for the reaction mechanisms of $\gamma p\rightarrow K^{\ast 0}\Sigma^+$, and provide further cue for the existence of the hidden-strange molecular states $N(2080){3/2}^-$and $N(2270){3/2}^-$.

In Fig.~\ref{fig:asy}, we show the theoretical results for the beam asymmetry ($\Sigma$), target asymmetry ($T$), and recoil asymmetry ($P$) predicted in the models of both the present work and Ref.~\cite{Wang:2018vlv}.
In Fig.~\ref{fig:asy}, the upper three panels and lower three panels show the corresponding results for the $\gamma p \rightarrow K^{\ast +}\Sigma^0$ and $\gamma p \rightarrow K^{\ast 0}\Sigma^+$ reactions, respectively. The blue solid lines and red dashed lines represent the corresponding results from the present work and Ref.~\cite{Wang:2018vlv}, respectively.
One sees that for both reactions, these spin observables calculated in the present work are quite different from those obtained in Ref.~\cite{Wang:2018vlv}.
We hope that these observables can be measured in the near future in experiments, as they can help to distinguish the models of the present work and Ref.~\cite{Wang:2018vlv}, and thus can provide further evidences for the existence of the molecular states $N(2080){3/2}^-$ and $N(2270){3/2}^-$.

From the results shown and discussed above, one sees that the available cross-section data for both $\gamma p \rightarrow K^{\ast +}\Sigma^0$ and $\gamma p \rightarrow K^{\ast 0}\Sigma^+$ in the near-threshold energy region can be well described in both the present work and Ref.~\cite{Wang:2018vlv}. However, the reaction mechanisms extracted from these two works are quite different. In particular, the resonance $\Delta(1905){5/2}^+$ introduced in Ref.~\cite{Wang:2018vlv} is now replaced in the present work by $N(2080){3/2}^-$, a $K^\ast\Sigma$ molecular state proposed in Refs.~\cite{He:2017aps,Lin:2018kcc} as strange partner of the $P_c(4457)$ state, and $N(2270)3/2^-$, a $K^\ast\Sigma^\ast$ molecular state as the strange partner of the $\bar{D}^\ast \Sigma^\ast_c$ bound states predicted as members in the same heavy-quark spin symmetry multiplet as the $P_c$ states \cite{Liu:2019}. Even though we cannot prefer one model against the other at the moment, it seems to be appropriate to say that the available cross-section data for $\gamma p \rightarrow K^{\ast +}\Sigma^0$ and $\gamma p \rightarrow K^{\ast 0}\Sigma^+$ do not exclude the possibility of the existence of the $N(2080){3/2}^-$ state as a $K^\ast\Sigma$ shallowly bound state and the $N(2270){3/2}^-$ state as a $K^\ast\Sigma^\ast$ shallowly bound state.

\section{Summary and Conclusion} \label{sec:summary}

In literature, one of the plausible explanations of the $P_c^+(4380)$ and $P_c^+(4457)$ states is that they are $\bar{D}\Sigma_c^\ast$ and $\bar{D}^\ast\Sigma_c$ molecules as their masses are just below the $\bar{D}\Sigma_c^\ast$ and $\bar{D}^\ast\Sigma_c$ thresholds. In this scenario, the $\bar{D}^\ast\Sigma_c^\ast$ bound states are also predicted to be in the same heavy-quark spin symmetry multiplet as the $P_c$ states. Analogously, in the light quark sector, the $N(1875){3/2}^-$ and $N(2080){3/2}^-$ states are proposed to be $K\Sigma^\ast$ and $K^\ast\Sigma$ molecules as strange partners of the $P_c^+(4380)$ and $P_c^+(4457)$ states \cite{He:2017aps,Lin:2018kcc}, and the $N(2270)$ state is proposed to be $K^\ast\Sigma^\ast$ molecule as strange partner of the $\bar{D}^\ast\Sigma_c^\ast$ bound states \cite{Liu:2019}. In the present work, we study the $\gamma p \rightarrow K^{\ast +}\Sigma^0$ and $\gamma p \rightarrow K^{\ast 0}\Sigma^+$ reactions to check if the $K^\ast \Sigma$ molecular picture of $N(2080){3/2}^-$ and  $K^\ast \Sigma^*$ molecular picture of $N(2270){3/2}^-$ are compatible with the available data for $K^\ast \Sigma$ photoproduction reactions.

The $\gamma p \rightarrow K^{\ast +}\Sigma^0$ and $\gamma p \rightarrow K^{\ast 0}\Sigma^+$ reactions have already been investigated in Ref.~\cite{Wang:2018vlv} within an effective Lagrangian approach. There, the $t$-channel $K$, $\kappa$, $K^\ast$ exchanges, the $s$-channel $N$, $\Delta$, $\Delta(1905){5/2}^+$ exchanges, the $u$-channel $\Lambda$, $\Sigma$, $\Sigma^\ast$ exchanges, and the generalized contact term were took into account in constructing the reaction amplitudes, and all the available data for both $\gamma p \rightarrow K^{\ast +}\Sigma^0$ and $\gamma p \rightarrow K^{\ast 0}\Sigma^+$ were well reproduced. It was found in Ref.~\cite{Wang:2018vlv} that the cross sections of $\gamma p \rightarrow K^{\ast +}\Sigma^0$ are dominated by the $s$-channel $\Delta(1905){5/2}^+$ exchange at low energies and $t$-channel $K^\ast$ exchange at high energies, with the $s$-channel $\Delta$ exchange providing significant contributions in the near-threshold region, and the cross sections of $\gamma p \rightarrow K^{\ast 0}\Sigma^+$ are dominated by the $t$-channel $K$ exchange at forward angles and $u$-channel $\Sigma^\ast$ exchange at backward angles, with the $s$-channel $\Delta$ and $\Delta(1905){5/2}^+$ exchanges making considerable contributions at low energies.

In the present work, we restudy the $\gamma p \rightarrow K^{\ast +}\Sigma^0$ and $\gamma p \rightarrow K^{\ast 0}\Sigma^+$ reactions by employing the same theoretical framework as Ref.~\cite{Wang:2018vlv} except that the $\Delta(1905){5/2}^+$ resonance introduced in Ref.~\cite{Wang:2018vlv} is now replaced by the $N(2080){3/2}^-$ and $N(2270){3/2}^-$ molecular states. Our results show that all the available cross-section data for both $\gamma p \rightarrow K^{\ast +}\Sigma^0$ and $\gamma p \rightarrow K^{\ast 0}\Sigma^+$ reactions can be well described. Further analysis shows that the cross sections of both $\gamma p \rightarrow K^{\ast +}\Sigma^0$ and $\gamma p \rightarrow K^{\ast 0}\Sigma^+$ are dominated by the $s$-channel $N(2080){3/2}^-$, $N(2270){3/2}^-$, and $N$ exchanges in the low energy region. The $u$-channel $\Sigma^\ast$ exchange provides significant contributions at backward angles in the high energy region. At the forward angles in the high energy region, it is the $t$-channel $K^\ast$ exchange and $K$ exchange that provides considerable contributions in $\gamma p \rightarrow K^{\ast +}\Sigma^0$ and $\gamma p \rightarrow K^{\ast 0}\Sigma^+$, respectively.

Both models in the present work and Ref.~\cite{Wang:2018vlv} describe the available cross-section data of $\gamma p \rightarrow K^{\ast +}\Sigma^0$ and $\gamma p \rightarrow K^{\ast 0}\Sigma^+$ quite well in all the energy region considered, but the reaction mechanisms extracted from these two models are quite different. At the moment we cannot prefer one model against the other. Even though, we conclude from the present work that the molecular pictures of the $N(2080){3/2}^-$ and $N(2270){3/2}^-$ states are compatible with the available cross-section data of the $\gamma p \rightarrow K^{\ast +}\Sigma^0$ and $\gamma p \rightarrow K^{\ast 0}\Sigma^+$ reactions. Near the threshold energy region, the total cross sections for $\gamma p \rightarrow K^{\ast 0}\Sigma^+$ predicted in the present work are bigger than those for $\gamma p \rightarrow K^{\ast +}\Sigma^0$. While in Ref.~\cite{Wang:2018vlv}, opposite pattern is observed. The single spin observables $\Sigma$, $T$, and $P$ for both $\gamma p \rightarrow K^{\ast +}\Sigma^0$ and $\gamma p \rightarrow K^{\ast 0}\Sigma^+$ predicted in models of the present work and Ref.~\cite{Wang:2018vlv} are also presented, and it is found that they all are strongly model dependent. We hope that these observables can be measured in the near future in experiments, which can be used to further constrain the reaction mechanisms of $\gamma p \rightarrow K^{\ast +}\Sigma^0$ and $\gamma p \rightarrow K^{\ast 0}\Sigma^+$ and, in particular, to further verify the molecular scenario of the $N(2080){3/2}^-$ and $N(2270){3/2}^-$ states.

\begin{acknowledgments}
This work is partially supported by the National Natural Science Foundation of China under Grants No.~12175240, No.~12147153, No.~12070131001, No.~11835015, and No.~12047503, the Fundamental Research Funds for the Central Universities, the China Postdoctoral Science Foundation under Grant No.~2021M693141, and the Grant of Chinese Academy of
Sciences (XDB34030000).
\end{acknowledgments}

\end{document}